%% file: main.tex
\documentclass[10pt]{article}
\usepackage{graphicx}

\def\Title#1{\begin{center} {\Large #1 } \end{center}}
\def\Author#1{\begin{center}{ \sc #1} \end{center}}
\def\Address#1{\begin{center}{ \it #1} \end{center}}

\newcommand\pubblock{\rightline{\begin{tabular}{l} Proceedings of the Fifth Annual LHCP\\ \pubnumber\\
         \pubdate  \end{tabular}}}

\newenvironment{Abstract}{\begin{quotation} \begin{center} 
             \large ABSTRACT \end{center}\bigskip 
      \begin{center}\begin{large}}{\end{large}\end{center} \end{quotation}}

\newenvironment{Presented}{\begin{quotation} \begin{center} 
             PRESENTED AT\end{center}\bigskip 
      \begin{center}\begin{large}}{\end{large}\end{center} \end{quotation}}

\input{econfmacros.tex}

\usepackage{color}

\usepackage{xspace}
\usepackage{comment}
\newcommand{\micron}{\ensuremath{\mu \mbox{m}}\xspace}

\newcommand\pubnumber{}

\newcommand\pubdate{\today}

\def\affiliation{
Laboratory for High-Energy Physics\\
EPFL, Lausanne, Switzerland\\
On behalf of the LHCb SciFi Tracker group}

\begin{document}

\large
\begin{titlepage}
\pubblock

\vfill
\Title{SciFi: A large Scintillating Fibre Tracker for LHCb}
\vfill

\Author{Plamen Hopchev}
\Address{\affiliation}
\vfill
\begin{Abstract}

The LHCb detector will be upgraded during the Long Shutdown 2 (LS2) of the LHC in order to cope
with higher instantaneous luminosities and to read out the data at 40~MHz using a trigger-less
readout system. The current LHCb main tracking system will be replaced by a single homogeneous
detector based on scintillating fibres. This contribution gives an overview of the
scintillating fibre tracker concept and presents the experience from the series production
complemented by most recent test-beam and laboratory results.

\end{Abstract}
\vfill

\begin{Presented}
The Fifth Annual Conference\\
 on Large Hadron Collider Physics \\
Shanghai Jiao Tong University, Shanghai, China\\ 
May 15-20, 2017
\end{Presented}
\vfill
\end{titlepage}
\def\thefootnote{\fnsymbol{footnote}}
\setcounter{footnote}{0}
%

\normalsize 


\section{Introduction}

The LHCb upgrade in Long Shutdown 2 of the LHC (2019--2020) will improve significantly
the experiment sensitivity in the flavour physics sector and will extend the LHCb physics
programme~\cite{bib_intro1,bib_intro2}. Along with sub-detector replacement and enhancements,
LHCb will employ a new fully software trigger with superior efficiency and flexibility.
The instantaneous luminosity will be increased by a factor of five to
$10^{33}~\rm{cm}^{-2}\rm{s}^{-1}$, aiming to collect 50~fb\textsuperscript{--1} over 10 years.

\begin{figure}[htb]
\centering
\includegraphics[width=0.8\textwidth]{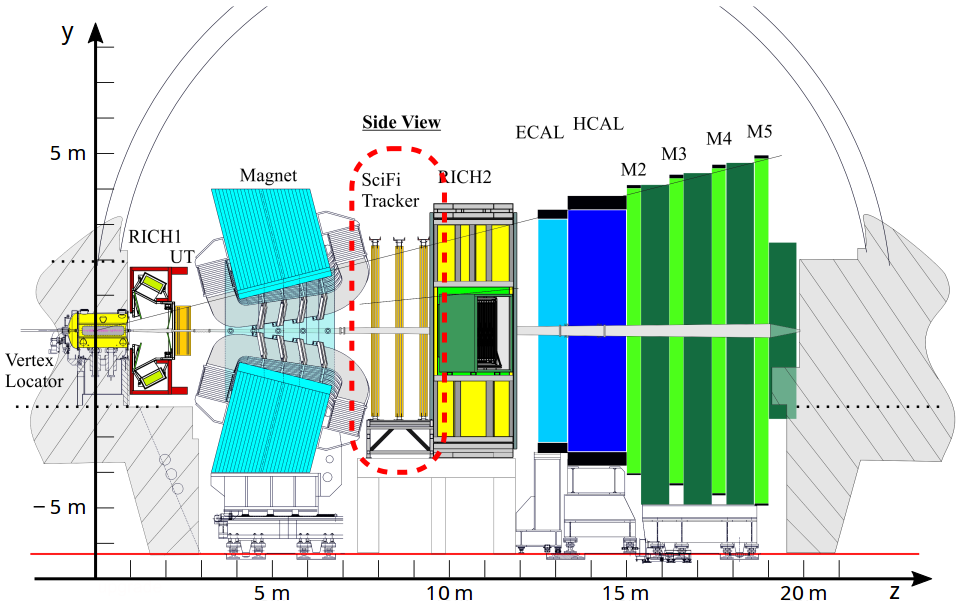}
\caption{Side view of the LHCb upgrade detector. The Scintillating Fibre Tracker will be
  installed in the tracking stations located downstream of the LHCb dipole magnet (highlighted in red).}
\label{fig:lhcb}
\end{figure}

The current LHCb main tracking system, composed of an inner and outer tracking
detector, will not be able to cope with the increased particle multiplicities and will be
replaced by a single homogeneous detector based on scintillating fibres (see
Fig.~\ref{fig:lhcb}). The new Scintillating Fibre (SciFi) Tracker will cover a total detector
area of 340~m\textsuperscript{2} and will provide a spatial resolution for charged particles
better than 100~\micron in the bending direction of the LHCb spectrometer~\cite{bib_scifi}.

\section{Scintillating Fibre Tracker}

\begin{figure}[htb]
\centering
\includegraphics[width=0.6\textwidth]{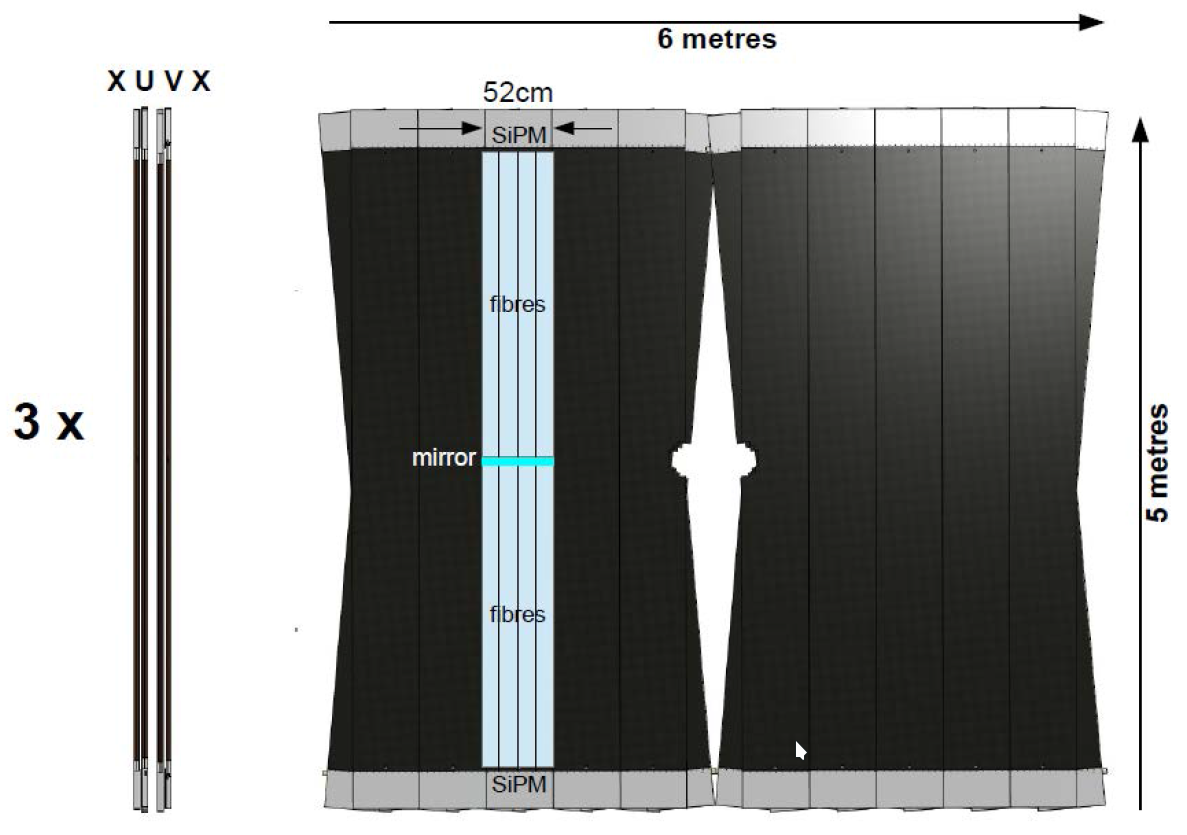}
\caption{Schematic yz- and xy-view of one SciFi tracking station. It is composed out of
  4 layers with vertical (x) and rotated (u,v) fibre orientations. Each layer is
  made of 10 or 12 individual fibre modules.}
\label{fig:scifi}
\end{figure}

The SciFi tracker consists of three stations each with four detection planes, as shown in
Fig.~\ref{fig:scifi}. The detector is built from individual modules (0.5~m $\times$ 4.8~m), each
comprising 8~fibre mats with a length of 2.4~m as active detector material. The fibre mats
consist of 6~layers of densely packed blue-emitting scintillating fibres with a diameter of
250~\micron. The scintillation light is recorded with arrays of state-of-the-art multi-channel
silicon photomultipliers (SiPMs). A custom ASIC is used to digitize the SiPM signals.
Subsequent digital electronics performs clustering and data-compression before the data is sent
via optical links to the DAQ system. To reduce the thermal noise of the SiPM, in particular
after being exposed to a neutron fluence of up to 10\textsuperscript{12}~neq/cm\textsuperscript{2}, 
expected for the lifetime of the detector, the SiPM arrays are mounted in so called cold-boxes and
cooled down by 3D-printed titanium cold-bars to $-40^\circ$~C.

The detector is designed to provide low material budget (1~\% per layer), hit efficiency
of 99~\% and a resolution better than 100~\micron. These performance figures must be
maintained over the lifetime of the detector which will receive radiation dose up to 35~kGy
near the beam pipe. The full detector, comprising 590$\,$000 channels, is read out at 40~MHz.  

The rest of this section provides information on the main detector components.

\subsection{Scintillating Fibres}

\begin{figure}[htb]
\centering
\includegraphics[width=1.0\textwidth]{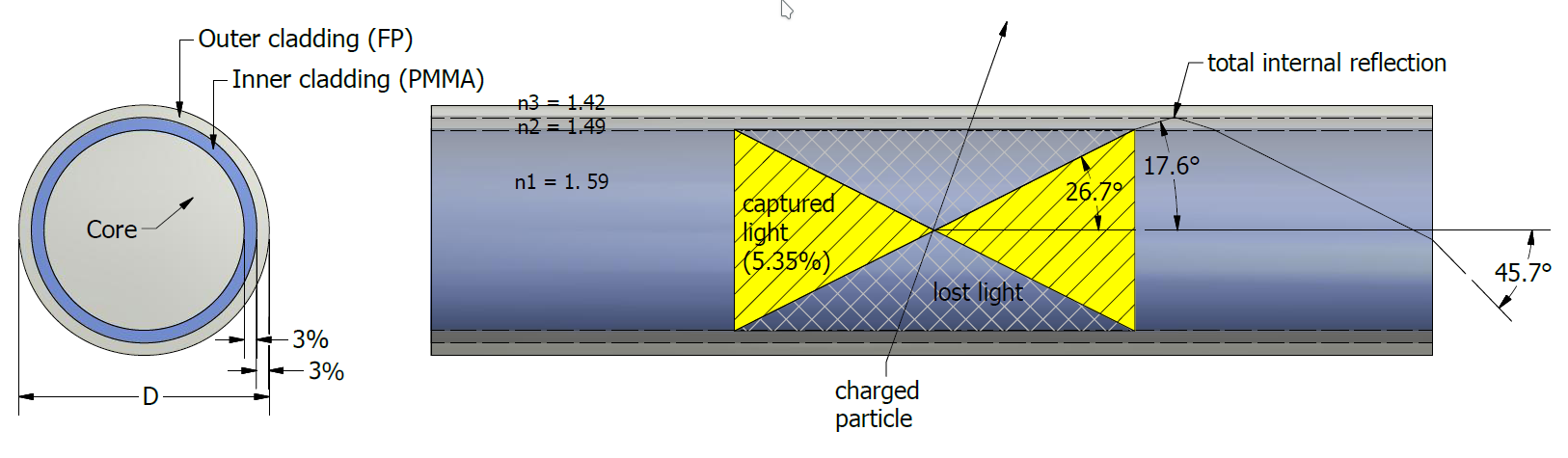}
\caption{Transverse (left) and longitudinal (right) section of a double cladded fibre,
  with a schematic representation of the light generation and transport.}
\label{fig:scint_fibre}
\end{figure}

The fibre type is SCSF-78MJ, produced by Kuraray, Japan. It has a diameter of 0.25~mm and is made of
polystyrene core with added dye and wavelength shifter, and two claddings with lower
refraction index (see Fig.~\ref{fig:scint_fibre}).

\begin{figure}[htb]
\centering
\includegraphics[width=1.0\textwidth]{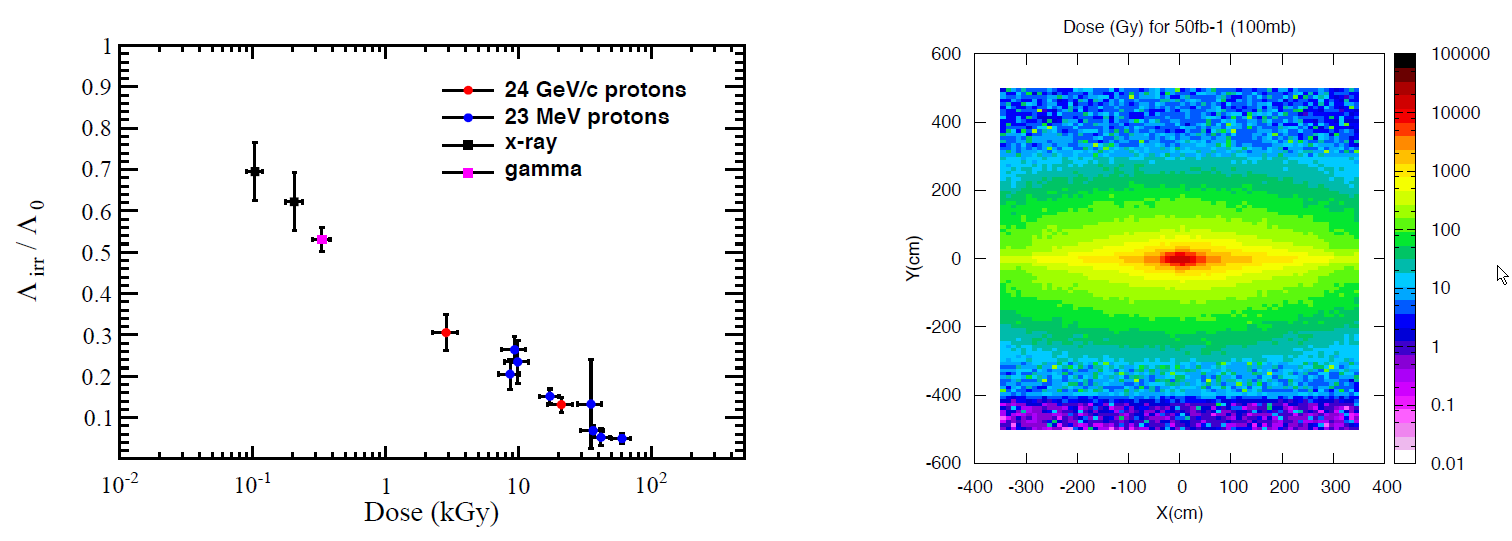}
\caption{Left: Reduction of the scintillating fibre attenuation length as function of
  accumulated ionizing dose for different particle types and energies. Right: Total
  expected ionzing dose at the location of the SciFi tracker, obtained from a FLUKA simulation.}
\label{fig:ALDose}
\end{figure}

The fibre produces several thousand photons per 1~MeV deposited energy, with a capture fraction
of about 5~\%~\cite{bib_scifi}. The decay time constant of the scintillation light signal is
2.8~ns, while the emission spectrum extends from about 400 to 600~nm and peaks at 450~nm near
the source. The attenuation length is greater than 3~m, with small dependence on the
wavelength. However, the ionizing radiation degrades the optical transparency of the fibre.
Dedicated irradiation tests with different particle types and energies were performed in order
to quantify this effect. It was estimated that at the end of the detector lifetime the signal
amplitude from the innermost part of the detector will be reduced by up to
40~\%~(dose 35~kGy)~\cite{bib_scifi}, while the radiation effect becomes relatively marginal
at distances of about 50~cm from the beam pipe (dose $<$ 1~kGy), see Fig.~\ref{fig:ALDose}.
The radiation effects were taken into account in the detector design. Replacement modules
are prepared for the areas exposed to highest radiation.

\subsection{Fibre mats and modules}

\begin{figure}[htb]
\centering
\includegraphics[width=0.6\textwidth]{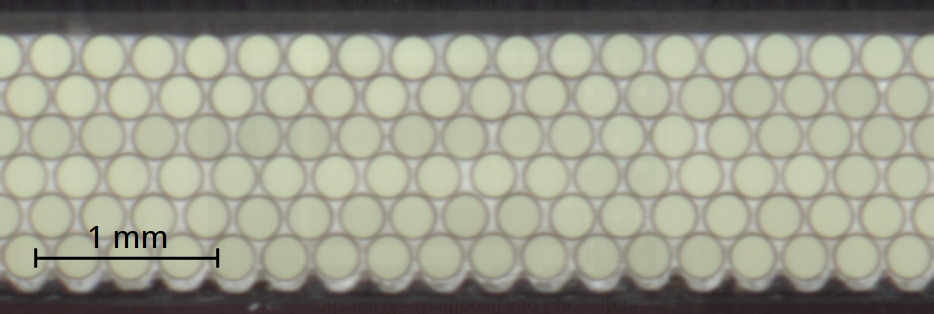}
\caption{Cross-section of a scintillating fibre mat.}
\label{fig:matscan}
\end{figure}

The active elements of the detector are scintillating fibre mats composed of six fibre
layers, with dimensions width $\times$ length $\times$ height: 130.65 $\times$ 2424.0
$\times$ 1.4~mm. A mat cross-section is shown in Fig.~\ref{fig:matscan}.

The mats are produced on a winding wheel with threaded surface with pitch of 275~\micron. Epoxy
glue is applied before each fibre layer and TiO$_2$ powder is added to the glue to reduce the optical
cross-talk between fibres. Reference alignment pins are produced on the mats from cavities on
the winding wheel surface. These alignment pins are used later during the module assembly to
ensure proper alignment of the mats. The scintillation light is detected only on one end of the
mat, and a mirror is glue on the other end to maximize the signal amplitude.

\begin{figure}[htb]
\centering
\includegraphics[width=0.65\textwidth]{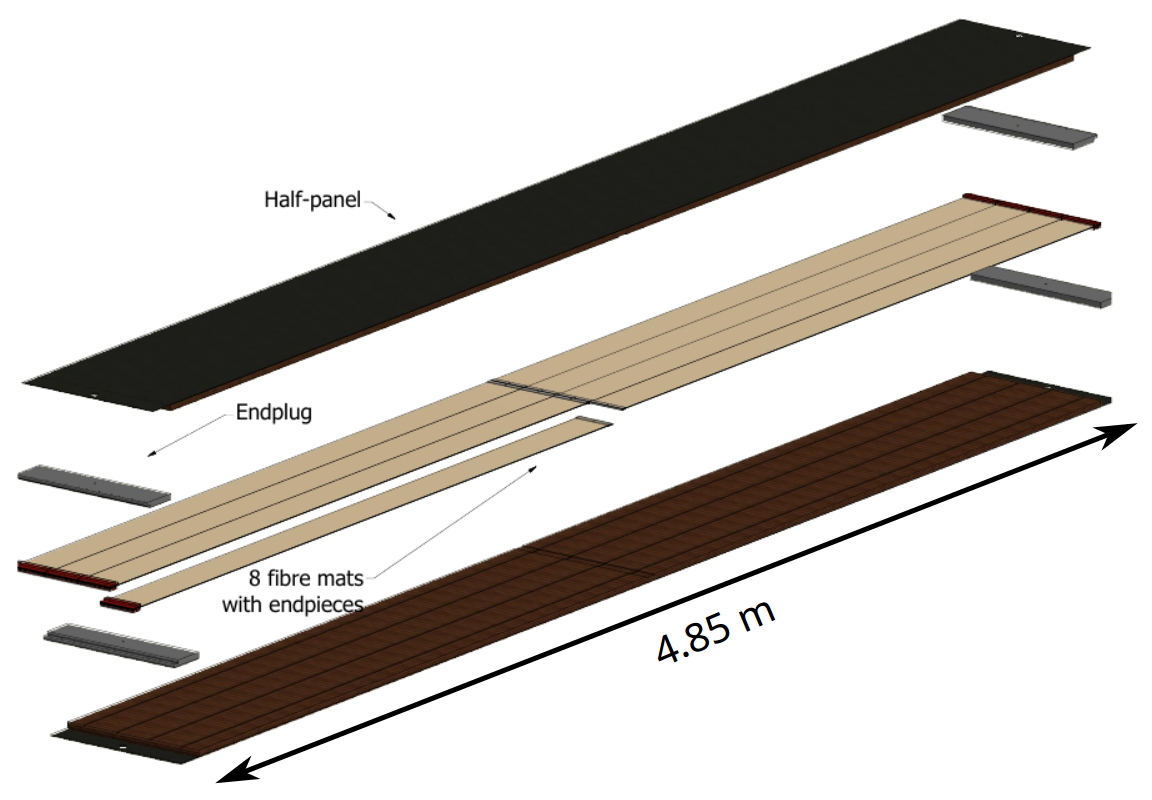}
\caption{A schematic of a SciFi module.}
\label{fig:module}
\end{figure}

During the module assembly 8 fibre mats are aligned and glued together with honeycomb/carbon-fibre
composite panels (see Fig.~\ref{fig:module}). Cooling enclosure, SiPMs and electronics are
added later. Despite the large dimensions (approx. $5 \times 0.5$~m) the construction
method results into stable and stiff detector modules with material budget of only 1.1\%~$X_0$
per module.

\subsection{Silicon Photomultipliers}

\begin{figure}[htb]
\centering
\includegraphics[width=0.9\textwidth]{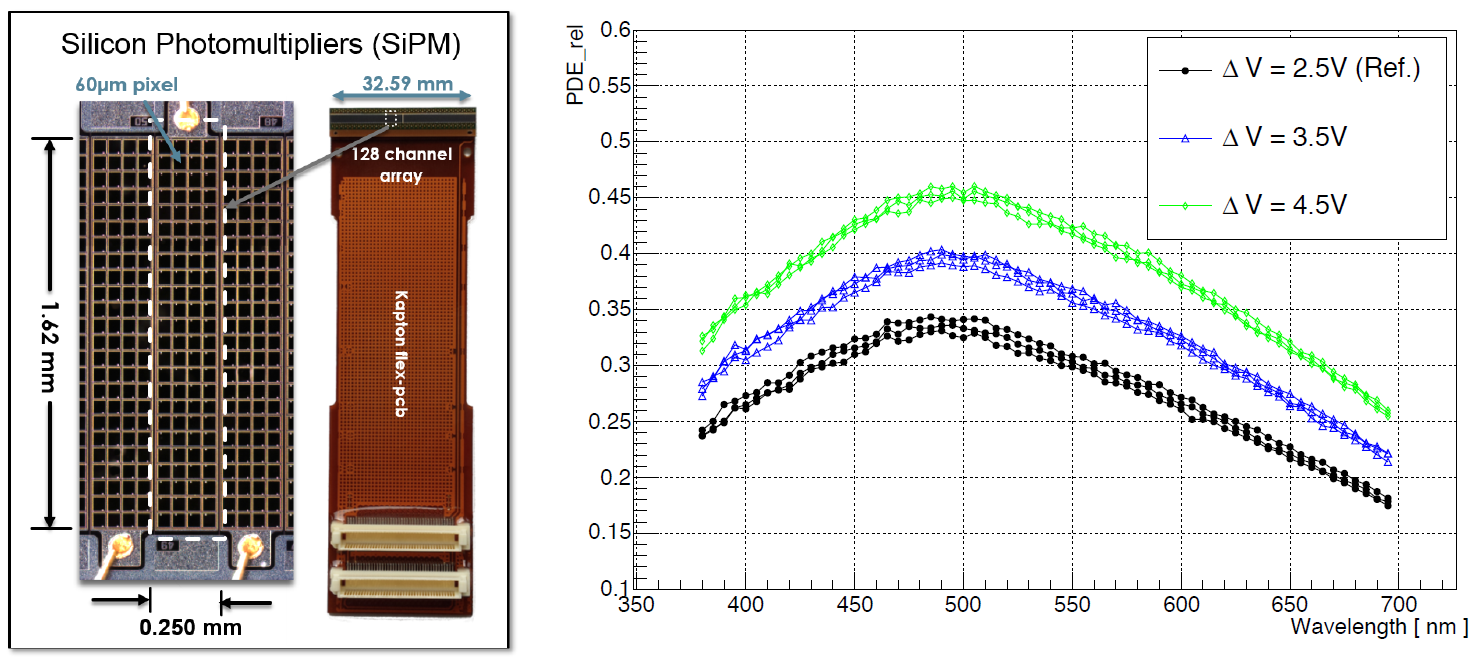}
\caption{Left: A close view on single SiPM channel, and a 128-channel SiPM array from Hamamatsu
  bonded to a flex-cable. Right: Photo detection efficiency as a function of wavelength measured at
  different over-voltage.}
\label{fig:sipm}
\end{figure}

The fibres are read-out by 128-channel silicon photo-multiplier (SiPM) arrays from
Hamamatsu, with individual channel size of 0.25 $\times$ 1.62~mm (see Fig.~\ref{fig:sipm}).
Several development iterations were made to improve the SiPM characteristics. The SiPM version
for the SciFi tracker provides photon detection efficiency about 45~\% at the nominal overvoltage 3.5~V,
and relatively low noise and cross-talk. To suppress dark noise after irradiation, the SiPMs
will be cooled to $-40^\circ$~C during operations. Irradiation tests showed that the SiPM dark
count rate (DCR) increases linearly with the neutron fluence. After receiving neutron fluence
of $6 \times 10^{11}$~neq/cm\textsuperscript{2}, which is about half of the expected fluence
for the lifetime of the detector, the SiPMs produce DCR of 14~MHz per channel at
$-40^\circ$~C. Such DCR rates pose no problem for the detector operation, because the detector
will be read out at 40~MHz, retaining only signal clusters of at least 4.5 photo-electrons.

\subsection{Front-end Electronics}

For efficient testing and maintenance, the front-end signal processing is performed on three
separate electronics boards. Initially, the SiPM signal is read by PACIFIC -- a custom
64-channel front-end ASIC developed for the SciFi tracker. The PACIFIC features a
pre-amplifier, a fast shaper, and dual gated integrators providing almost zero dead-time.
The analog signal is digitized by a triple-threshold tunable comparator. The digitized channel
data are sent to a clusterization board where FPGAs combine the individual channels into clusters.
Later on, a master board sends the data via optical links to the back-end DAQ system.

\section{Testbeam performance}

\begin{figure}[htb]
\centering
\includegraphics[height=0.23\textheight]{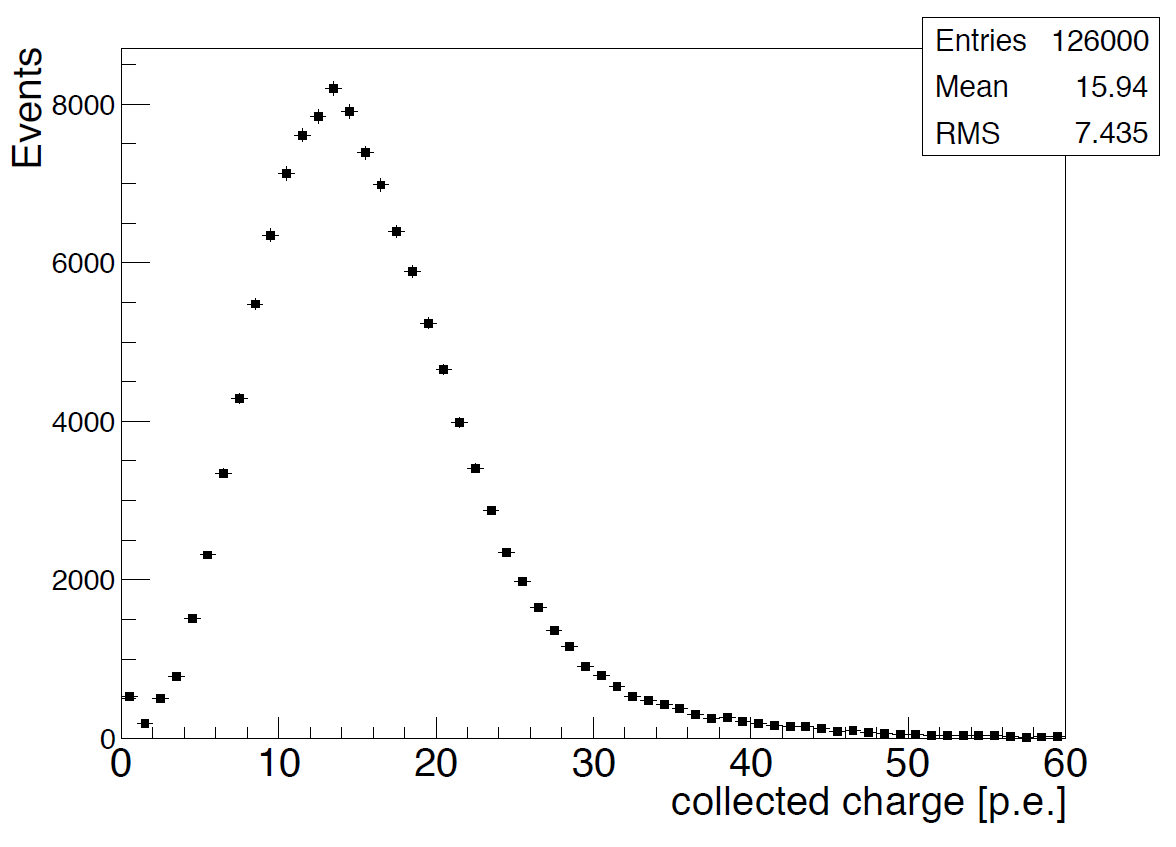}
\hspace{3ex}
\includegraphics[height=0.22\textheight]{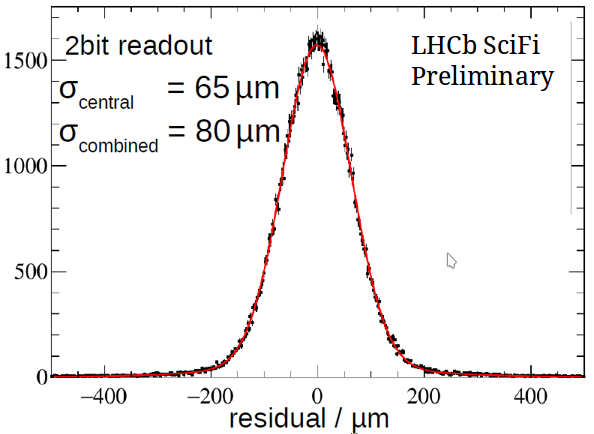}
\caption{Testbeam results at CERN and DESY. Left: Light yield measured close to the
  mirror. Right: hit resolution using SciFi readout.}
\label{fig:testbeam}
\end{figure}

Several testbeam campaigns demonstrated good performance of the SciFi mats and
electronics. The main detector characteristics were measured with proton/pion beams
at CERN and electron beams at DESY, using the nominal SciFi readout components. Nominal light
yield of 16 photo-electrons was achieved, as well as 99~\% hit efficiency and 70~\micron
hit resolution (see Fig.~\ref{fig:testbeam}). Fibre mats and SiPMs irradiated in different
scenarios and dose were measured too, showing the expected behavior.

\section{Conclusion}

The LHCb Upgrade Scintillating Fibre tracker is a high-resolution detector covering area of
340~m$^2$. It is based on scintillating fibres with diameter 0.25 mm, read-out with silicon
photomultipliers. Nominal performance parameters have been achieved in laboratory and testbeam
measurements. The 2.4-m long fibre mats provide light yield $\geq$ 16 photo-electrons,
99~\% hit efficiency and position resolution of 70~\micron. The serial production of the detector
components is progressing well, and the detector installation is foreseen to start in 2019.




\end{document}

%% file: econfmacros.tex
\def\beq{\begin{equation}}
\def\eeq#1{\label{#1}\end{equation}}
\def\eeqn{\end{equation}}

\def\beqa{\begin{eqnarray}}
\def\eeqa#1{\label{#1}\end{eqnarray}}
\def\eeqan{\end{eqnarray}}

\let\bar=\overbar

\def\Dslash{\not{\hbox{\kern-4pt $D$}}}
\def\dslash{\not{\hbox{\kern-2pt $\del$}}}

\def\msb{{\bar{\ssstyle M \kern -1pt S}}}

